\documentclass[pra,twocolumn,showpacs,superscriptaddress,preprintnumbers,amsmath,amssymb,aps]{revtex4-1}
\usepackage{mathrsfs}
\usepackage{amsmath}
\usepackage{amssymb}
\usepackage{graphicx}
\usepackage{dcolumn}
\usepackage{bm}
\usepackage{subfigure}
\usepackage{color}
\usepackage{ifpdf}
\usepackage{epstopdf}
\usepackage{CJK}
\usepackage{verbatim}

\usepackage[
pdfstartview=FitH,%
bookmarks=true,%
bookmarksopen=true,%
breaklinks=true,%
colorlinks=true,%
linkcolor=blue,anchorcolor=blue,%
citecolor=blue,filecolor=blue,%
menucolor=blue,pagecolor=blue,%
urlcolor=blue]{hyperref}
\usepackage{float}
\usepackage[toc,page,title,titletoc,header]{appendix}

\begin{document}
	%\begin{CJK*}{GBK}{song}
	%\begin{CJK*}{UTF8}{gbsn}
	
	\title{Parallel multicomponent interferometer with a spinor Bose-Einstein condensate}
	\author{Pengju Tang}
	\affiliation{State Key Laboratory of Advanced Optical Communication System and Network, Department of Electronics, Peking University, Beijing 100871, China}
	\author{Peng Peng}
	\affiliation{State Key Laboratory of Advanced Optical Communication System and Network, Department of Electronics, Peking University, Beijing 100871, China}
	\author{Zhihan Li}
	\affiliation{State Key Laboratory of Advanced Optical Communication System and Network, Department of Electronics, Peking University, Beijing 100871, China}
	\author{Xuzong Chen}
	\affiliation{State Key Laboratory of Advanced Optical Communication System and Network, Department of Electronics, Peking University, Beijing 100871, China}
	\author{Xiaopeng Li}\email{xiaopeng\_li@fudan.edu.cn}
	\affiliation{State Key Laboratory of Surface Physics, Institute of Nano-electronics and Quantum Computing, and Department of Physics, Fudan University, Shanghai 200433, China}
	\affiliation{Collaborative Innovation Center of Advanced Microstructures, Nanjing 210093, China}
	\author{Xiaoji Zhou}\email{xjzhou@pku.edu.cn}
	\affiliation{State Key Laboratory of Advanced Optical Communication System and Network, Department of Electronics, Peking University, Beijing 100871, China}
	\affiliation{Collaborative Innovation Center of Extreme Optics, Shanxi University, Taiyuan, Shanxi 030006, China}
	
	\date{\today}
	
\begin{abstract}
Atom interferometry with high visibility is of high demand for precision measurements. Here, a parallel multicomponent interferometer is achieved by preparing a spin-$2$ Bose-Einstein condensate of $^{87}$Rb atoms confined in a hybrid magneto-optical trap. After the preparation of a spinor Bose-Einstein condensate with spin degrees of freedom entangled, we observe four spatial interference patterns in each run of measurements corresponding to four hyperfine states we mainly populate in the experiment. The atomic populations in different Zeeman sublevels are made controllably using magnetic-field-pulse induced Majorana transitions. The spatial separation of atom cloud in different hyperfine states is reached by Stern-Gerlach momentum splitting. The high visibility of the interference fringes is reached by designing a proper overlap of the interfering wave packets. Due to uncontrollable phase accumulation in Majorana transitions, the phase of each individual spin is found to be subjected to unreproducible shift in multiple experimental runs. However, the relative phase across different spins is stable, paving a way towards noise-resilient multicomponent parallel interferometers.
\end{abstract}
	
\pacs{32.80.Qk,03.75.Dg,02.30.Yy,03.75.-b}
	
\maketitle

\section{Introduction}
Quantum mechanical particles such as photons, electrons, and atoms have been used to construct two-slit interferometry, which plays important roles in studying fundamental quantum theories and enables high-sensitivity measurements, leading to important applications such as quantum precision measurement, quantum information and quantum simulation~\cite{39RevModPhys,5Stamper-Kurn2013,43Negretti2011,29PRLinterferency,47shortcut}. For example, the ``smokingkun" experimental demonstration of phase coherence in a Bose-Einstein condensate (BEC) has been achieved with atom interferometry~\cite{31Andrews637}. Matter-wave interferometers with long coherence time using ultracold atom gases can be used for high-precision measurements in a broad context, e.g., in the study of the quantum properties of atoms~\cite{1KaiBongs2004} and the quantum phase evolution of the wave function~\cite{13Simsarian2000,38Matter,40wheeler2004spontaneous,24PhysRevA}, the correlation characteristics~\cite{20Polkovnikov2006}, the dynamics of isolated quantum many-body systems \cite{34Gring1318,35correlations,36Langen207,44RevModPhys.83.863,44Thermalization}, and gravitational effects~\cite{41Dimopoulos2007Relativity,42HMuntinga2013Relativity}.

Recently, there have been growing efforts in developing matter wave interferometry with unconventional approaches. Ramsey interferometers (RIs) with atomic external motion states of a BEC trapped in a harmonic potential have been demonstrated with high interferometric visibility for several cycles~\cite{45Interferometry,26Ramsey}. Coherent superposition of different momentum-spin states that entangle internal and external states has been achieved~\cite{11Machluf2013}. The interferometer based on the spin entanglement can be used to study quantum effects of gravity~\cite{32PhysRevLett} and construct quantum simulation platforms. Atomic clock interferometer has offered a promising high-precision tool to study the interplay of general relativity and quantum physics, which can help to formulate a modern version of the uncertainty principle in terms of entropies and deepen our understanding of the wave-particle duality~\cite{33Zych2011clock,margalit2015self,27Gravity,28Pikovski2017clock}.
	
In the applications of atom interferometers, it is crucial to further improve the sensitivity and reduce the apparatus complexity, which has motivated the development of a multistate interferometric scheme~\cite{3petrovic2013}. The spatial interference fringes would provide a solid experimental evidence for phase coherence~\cite{9Miller2005,10Yinbiao2015}. However, these experimental atomic spatial interference fringes have been rarely observed for high spin atom systems, although this represents one promising experimental platform to implement multistate interferometers. The contrast and phase stability are important parameters of atom interferometers and are important for studying coherence and correlation properties~\cite{15Rath2010,17Watanabe2012,37PhysRevA}.

Here we report the realization of a parallel multicomponent interferometer, utilizing the BEC phase coherence in a spin-2 cold atom gas. In our experiment, interference fringes have been observed in different spin channels. By properly optimizing experimental parameters, a high-visibility multispin parallel interferometer is achieved. By extracting the information of the interference fringes, we find that the phase of a single component is unstable, i.e., fluctuates in repeated experiments, but the relative phase of the extracted different components remains stable. This achieved multicomponent spatial interferometer can be a potential application for high-precision measurement, such as measuring the small changes in the spin-dependent external potential~\cite{39RevModPhys} or the phase profile of an evolving multispin BEC~\cite{13Simsarian2000,5Stamper-Kurn2013}.

\section{Experimental demonstration and data analysis}
\begin{figure}[htp]
\includegraphics[width=\linewidth]{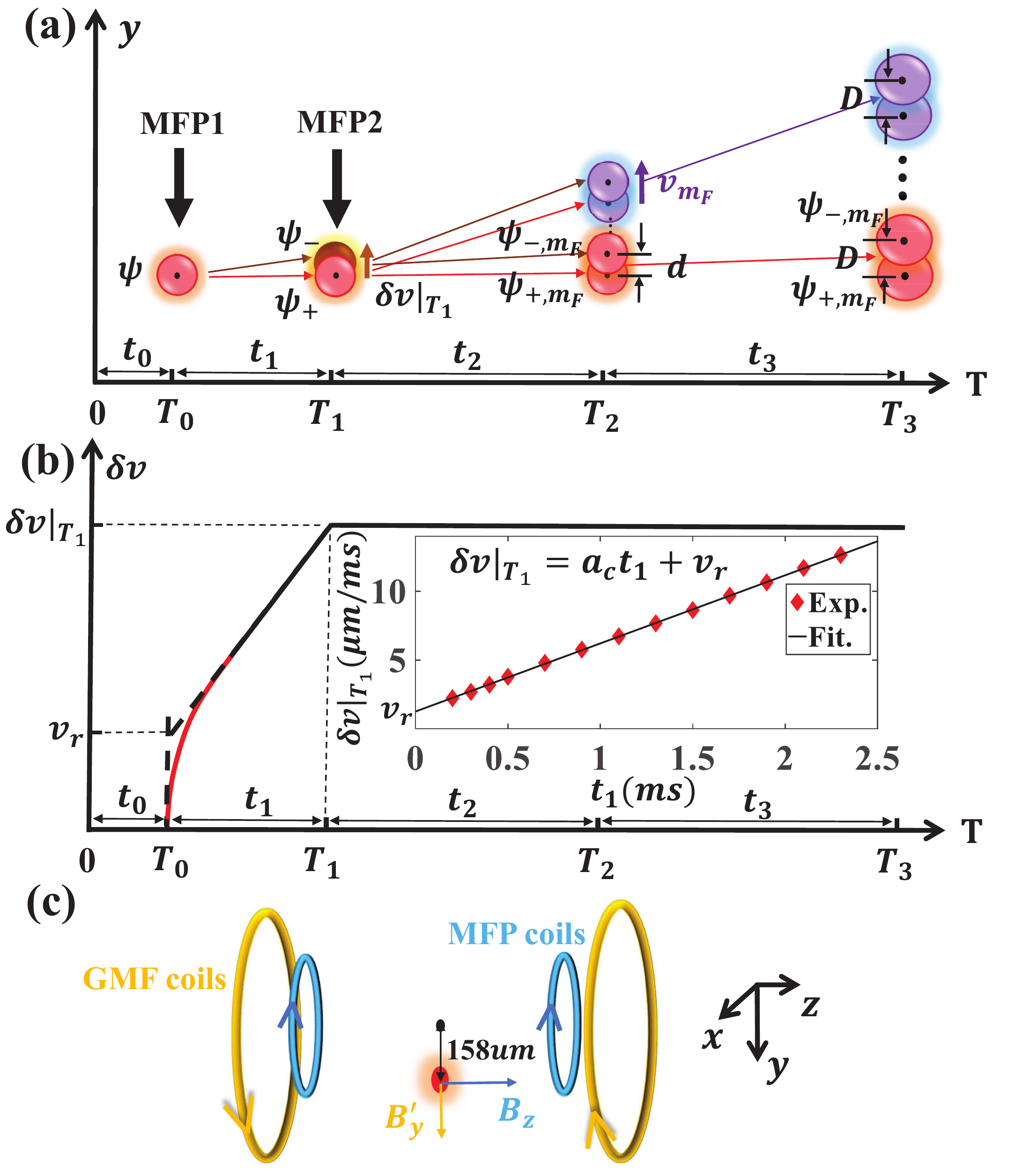}
\caption{(Color online) (a) Space-time diagram of the parallel multicomponent interferometer with a spinor BEC. Two Majorana transitions achieved by precisely controlling the falling curve of the MFP1 and MFP2 and the two Stern-Gerlach processes induced by the GMF during the interval $t_1$ and $t_2$ form the interferometer along $y$ axis (the gravity direction). (b) The schematic diagram of velocity difference $\delta v$ of the interfering wave packets at each stage of the interferometer. At the beginning, the red solid line indicates the $\delta v$ mainly comes from the repulsive interaction between the wave packets $\psi_{+}$ and  $\psi_{-}$, which contributed a velocity $v_r$ as shown by the intersection of two tangent dashed lines of the red curve.  During the time $t_1$, the black solid line and  its extended dotted line represent the velocity difference between $\psi_{+}$ and  $\psi_{-}$ caused by different external potentials. The velocity difference remains a constant $\delta v|_{T_1}$ after time $T_{1}$, which refers to the velocity difference between $\psi_{+,m_F}$ and $\psi_{-,m_F}$. The inset shows $\delta v|_{T_1}$ as the function of $t_1$ after MFP1, where the experimental points of $\delta v|_{T_1}$ are equal to the vertical distance between $\psi_+$ and $\psi_-$ divided by $26 ms$ TOF. The points are fitted with $\delta v|_{T_1}={a_c}{t_1}+v_r$. (c)Experimental structure of the magnetic field coils. The GMF coils keeping open until the time $T_2$ supply a gradient field $B_y^{'}$ and the uniform magnetic field $B_z$ supplied by the MFP coils can be finely controlled  to achieve Majorana transition.}
\label{fig:procedure}
\end{figure}

The process of designing the parallel multicomponent atom interferometer with different submagnetic level $m_F$ and atomic spin $F$ can be seen in Fig.~\ref{fig:procedure}(a), where $y$ axis is along the direction of gravity. First, we prepare a condensate and let it moderately expand a period of time $t_0$. At the time $T_0$, we have the initial prepared condensate in the state $\left|F,m_F = F \right>$ with the wave packet $\psi$. Then it is projected into two submagnetic states $\left|F,m_F = F\right>$ and $\left|F,m_F =F-1 \right>$ corresponding to $\psi_+$ and $\psi_-$ after an optimized Majorana transition achieved by the first magnetic field pulse (MFP1). These condensates of the two states are allowed to evolve for a period of time $t_1$ in a gradient magnetic field (GMF).  After the first Stern-Gerlach process, the two wave packets ($\psi_+$ and $\psi_-$) gain a velocity difference $\delta v|_{T_1}$ at time $T_1$ as shown in Fig.~\ref{fig:procedure}(b), while they are still overlapping. Meanwhile, each of the two wave packets is converted into multiple $|m_F\rangle$ states (corresponding to $\psi_{+,m_F}$ and $\psi_{-,m_F}$) by another Majorana transition achieved by the second magnetic field pulse (MFP2). All these states are alllowed to evolve for a period of time $t_2$ in a GMF, which is the second Stern-Gerlach process. Then the center of mass of the combined wave packet ($\psi_{+,m_F}$ and $\psi_{-,m_F}$ with the separation $d$) has a velocity $v_{m_F}$ at time $T_2$. Finally, switching off all the trap and after a time of flight (TOF) $t_3$, the wave packets of the same submagnetic states with the separation $D$ will give an interference pattern at time $T_3$ and the different components will be separated in space due to the different velocity $v_{m_F}$. A parallel multicomponent interferometry is achieved in camera with the absorption imaging method. 
	
\subsection{Experimental implementation}
In the experiment, we prepare a BEC of $^{87}$Rb in an atomic hyperfine state $\left| {F,{m_F}} \right\rangle = \left| {2,2} \right\rangle $ in a hybrid trap, where the atomic spin $F=2$ and submagnetic level $m_F=2$, combining an optical dipole trap formed by a laser beam with a wavelength $1064$nm and a quadrupole magnetic trap with a field gradient $B_y^{'}=12.4$G/cm. The atom number is about $2.0 \times10^5$ and the temperature is $70$nK. The trapping frequencies in three directions are $(\omega_x,\omega_y,\omega_z)=2\pi\times$(28Hz, 55Hz, 65Hz)~\cite{25PhysRevA}, corresponding to the Thomas-Fermi diameter~\cite{4Dalfovo1999} of the condensate $25$$\mu$m, $8.6\mu$m, and $6.6 \mu$m, respectively.
	
After the preparation of BEC, we turn off the optical trap but keep the quadrupole magnetic trap open, then let the condensate expand a short time $t_0$ to make a bigger size, so as to have a better wave packet overlap in the interference. For reasons that the interval time $t_0$ between the shutoff of the optical trap and the MFP1 is short-around a few milliseconds, the Majorana loss can be negligible, during which the BEC moved only a few microns. At time $T_0$ the BEC is located about $158 \mu$m below the center of the quadrupole magnetic trap in $y$ direction, as is shown in Fig.~\ref{fig:procedure}(c). The magnetic field strength here is very small, so that we can control the transition of atoms in different submagnetic levels by the rotation of the magnetic field here~\cite{21Xiuquan2006}.

In order to make the atoms in different submagnetic states finely tunable, here we use the method of Majorana transition. Although it is a very old problem~\cite{46Majorana,23XiaLin2008}, the technical difficulty of using Majorana transition to construct the multicomponent interferometer lies in the precise control of spin flips at the time of applying the magnetic field pulse (MFP). The MFP consists of three processes: turn on adiabatically, keep a short time, and switch off nonadiabatically the magnetic field $B_z$ along the $z$ direction, which is achieved through finely controlling the current in MFP coils by a signal generator. Here the fine tunable spin projection in the Majorana transition is achieved by controlling shutoff time and curve of the MFP. The atomic Larmor frequency is about $274$kHz under the weak magnetic field along the $y$ direction supplied by GMF coils. The rising time of the MFP is chosen as $40 \mu$s, which is rather slow comparable to the Larmor frequency, so that the turn-on process is adiabatic.  The switch-off time can be as short as $5\mu$s, which is comparable to the Larmor precession and is thus nonadiabaticaly leading to Majorana transitions. In between the turning on and off of the pulse is kept by $20\mu$s to balance the magnetic field along $z$ axis, during which the magnetic field $B_z$ is about a few hundred milligauss supplied by the MFP coils with 1.1A current, as is shown in Fig.~\ref{fig:procedure}(c).
	
In order to project the BEC into two hyperfine states with approximately equal population at time $T_0$, the ramp-off time of MFP1 is $16\mu$s, where the falling curve consists of two straight lines with different slopes. Our measurement shows the resultant populations for the hyperfine states $\left|m_F\right\rangle = \left|2\right\rangle$(corresponding to $\psi_+$), $\left|m_F\right\rangle = \left|1\right\rangle$ (corresponding to $\psi_-$), and $\left|m_F\right\rangle=\left|0\right\rangle$ are $50\%$, $45\%$, and $5\%$, respectively. The populations for other states are essentially negligible. However, in order to populate the atoms in $\psi_+$ and $\psi_-$ into all hyperfine states at time $T_1$, the MFP2 is chosen to be identical to the MFP1 except that the ramp-off time is 8.5 $\mu$s, i.e. roughly two times faster than the ramp-off time of the MFP1. The detailed calculation of MFP2's role is given in Appendix~\ref{sec:Appendix1}.
	
During the interval $t_1$ the accumulated velocity difference $\delta v$ between $\psi_+$ and $\psi_-$ mainly comes from two different effects: the different action of the external potential (both magnetic field and gravity)~\cite{12Fort2001} on the wave packets $\psi_{+}$ and $\psi_{-}$, and the mutual repulsion between the two wave packets~\cite{13Simsarian2000,50machluf}. Due to the discrepancy of the external potential for the two wave packets, the accumulated velocity difference is ${a_c}{t_1}$, where $a_c$ is relative acceleration between adjacent submagnetic states. We measured it as $a_c=4.7\mu m/ms^2$, which is consistent with theoretical calculation (see more details in Appendix~\ref{sec:Appendix0}). The mutual repulsion between the two wave packets gives rise to a velocity $v_r=1.2\mu m/ms$, as shown in the inset of Fig.~\ref{fig:procedure}(b), which is consistent with the estimation using a scattering length $a_{s}=5.18n$m~\cite{8Boesten1997}. So $\delta v|_{T_1}$ is equal to ${a_c}{t_1}+v_r$ at time $T_1$.	
	
During the $t_2$ process the separation of the wave packets $\psi_{+,m_F}$ and $\psi_{-,m_F}$ keeps increasing, which is owing to their velocity difference $\delta v|_{T_1}$ as is shown in Fig.~\ref{fig:procedure}(b). The small changes (which are estimated to be less than $5\%$) in velocity difference between $\psi_{+,m_F}$ and $\psi_{-,m_F}$ can be ignored after the MFP1, which comes from the tiny position discrepancy of the two wave packets in the harmonic trap. So we have $\delta v|_{T_2} \thickapprox \delta v|_{T_1}$ for every hyperfine state. The eventual separation of the wave packets $\psi_{+,m_F}$ and $\psi_{-,m_F}$  at time $T_2 $ is $d \thickapprox 1/2{a_c}{t_1}^2+{v_r}{t_1}+\delta v|_{T_1} t_2$. Meanwhile, the center of mass of the combined wave packet ($\psi_{+,m_F}$ and $\psi_{-,m_F}$) acquires a velocity $ v_{m_F}\thickapprox(2-m_{F}){a_c}{t_2}$ due to the different action of the external potential on the wave packets of different $\left| m_F\right\rangle$ states during the $t_2$ process. In experiment, the velocity difference $v_{m_F}-v_{m_F-1}$ between adjacent $\left| m_F\right\rangle$ states is designed to be much greater than $\delta v|_{T_2}$, so that the wave packets of different components can be spatially separated after TOF.

\begin{figure}
	\includegraphics[width=\linewidth]{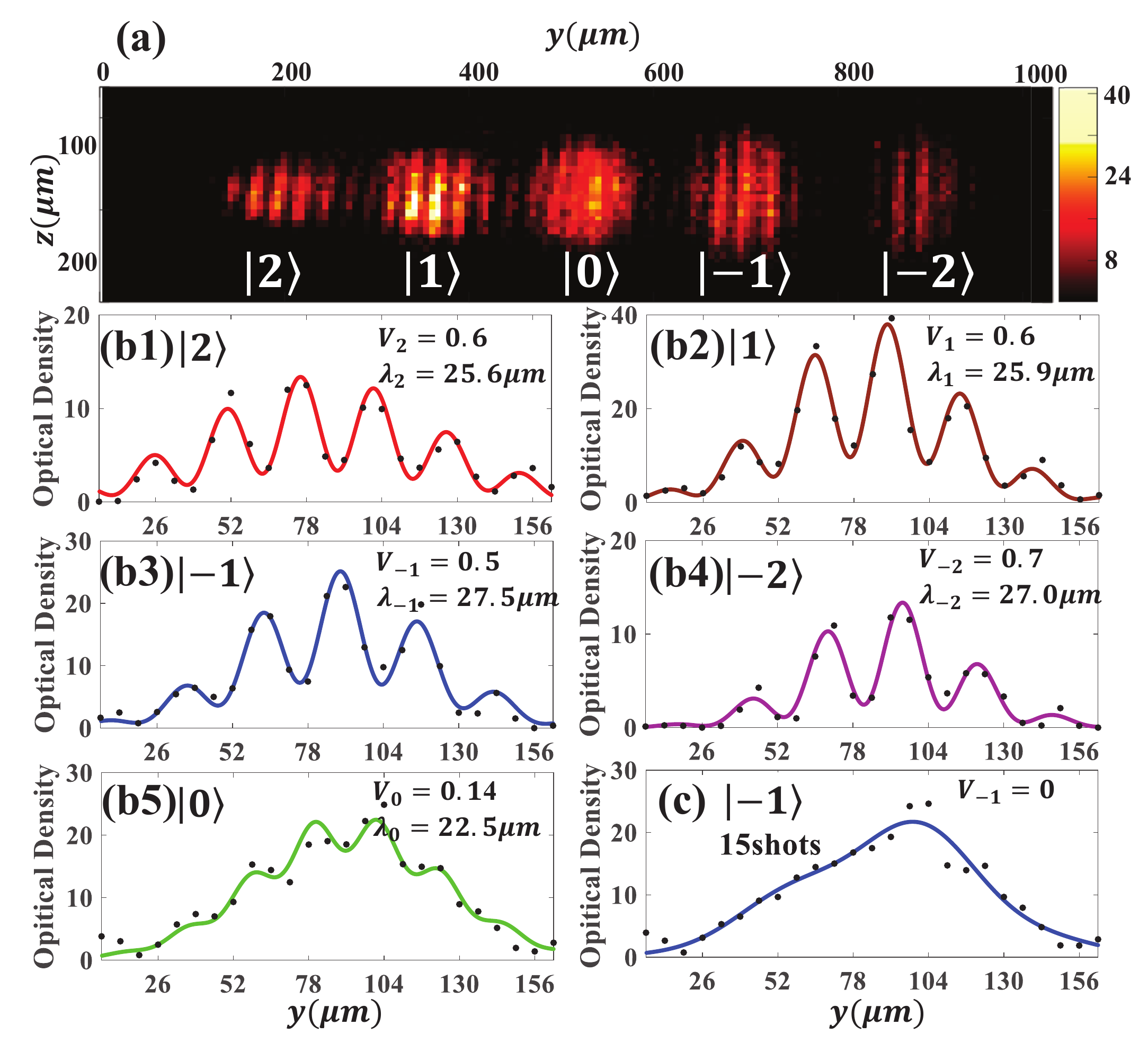}
	\caption{(Color online) (a) Typical interference picture in the $(y,z)$ plane. (b1)-(b5) The density distributions of five component interference patterns along $y$ axis, after integrating the patterns along $z$ axis in (a). The points are the experimental data and the curves are fitting results using Eq.~\eqref{eqn:fitting}. (c) The average of 15 consecutive experimental shots with a visibility reduction to zero for the chosen state $|m_{F}=-1\rangle$.}
	\label{fig:interference}
\end{figure}

Finally, we turn off the quadrupole magnetic trap, and all the wave packets will fall freely and expand. The same hyperfine state atoms would interfere in the ballistic expansion, where the separation of their wave packet centers is $D=d+\delta v|_{T_2} t_3$. A typical picture in the plane $(y, z)$ after $26 ms$ TOF is shown in Fig.~\ref{fig:interference}(a), where multiple interference fringes spatially separated corresponding to the five hyperfine components are observed with the chosen $t_1 = 210 \mu$s and $t_2 = 1300 \mu$s, where the time duration $t_1$ includes the rising and keeping time of MFP2. To show it more clearly, we integrate over the $z$ direction, and the density distributions along $y$ axis are given as the points in Figs.~\ref{fig:interference}($b1$)-~\ref{fig:interference}($b5$), corresponding to the different states $|2\rangle$, $|1\rangle$, $|-1\rangle$, $|-2\rangle$, and $|0\rangle$, respectively, where the interference patterns are clearly revealed. From these patterns, the interference information including the fringe frequency, visibility, and phase can be obtained.
	
\subsection{Theory of the multicomponent interferometer}
In order to understand the parallel multicomponent interference, we give a theory analysis for the interference pattern. The condensate wave function at time $T_3$ takes an approximate form with the wave packet width $\sigma_{m_F}$ in $y$ direction as~\cite{13Simsarian2000,11Machluf2013}
\begin {eqnarray}\label{eqn:wave1}	&&\psi_{q,\,m_{F}}(y) \\
&=&\Gamma_{q, m_F} \exp [-\alpha_{m_{F}}(y+q\frac{D}{2})^{2}+i\gamma_{q,m_{F}}(y+q\frac{D}{2}) + i\theta_{q, m_F}] \nonumber
\end {eqnarray}
which is associated with the hyperfine $|m_F \rangle $ component generated from the $q$th copy---$q = \pm$ labels the two copies of wave packets before applying the MFP2 [see Fig.~\ref{fig:procedure}(a)] $\Gamma_{q, m_F} $ represents the amplitude of the wave packet. The parameter $\alpha_{m_F}={1}/{2\sigma_{m_F}^2}-i{\beta_{m_F}}/{2}$ is a complex number, where the real part contains the wave packet width $\sigma_{m_F}=\sigma_{m_F}(0)\sqrt{1+t_3/\varepsilon_{m_F}}$ of the $|m_F\rangle$-state condensate after TOF and the imaginary part $\beta_{m_F}=M/\hbar(t_3+\varepsilon_{m_F})$ (with  $\varepsilon_{m_F}={M^2{\sigma }_{{m}_{F}}^{4}\left(0\right)}/{{\hslash }^{2}{t}_{3}}$) is the coefficient of the quadratic phase due to the evolution of the wave packets during the TOF stage~\cite{margalit2015self}. $\sigma_{m_F}(0)$ is the wave packet width of $|m_F\rangle$-state condensate in the trap, $\hbar$ is reduced Planck constant, and $M$ is the atomic mass. The coefficient $\gamma_{q,m_{F}}={M{v}_{q,m_{F}}}/{\hbar}$ of the linear term  in $y$ takes into account the center-of-mass motion, where $v_{q, m_F}$ is the propagation velocity of the center of mass of the wave packet $\psi_{q,m_F}$ and $v_{-, m_F}-v_{+, m_F}=\delta v|_{T_2}$. $\theta_{q, m_F}$ is the phase of the wave packet $\psi_{q,m_F}$ at the center.
	
The coherent superposition $|\psi_{+, m_F}(y)+\psi _{-,m_F}(y)|^{2}$ describes the interference patterns formed by the two copies of condensate populating in the $|m_F\rangle$ state. To exhibit the character of the interference pattern, it is sufficient that we consider the case $t_3\gg \varepsilon_{m_F}$ (the long TOF limit) and the final velocity difference between the two interfering wave packets is smaller than the expansion velocity of each one of them, in which the interfering wave packets are well overlapped~\cite{11Machluf2013,margalit2015self}. Then $\alpha_{m_F}$ is approximately given by $\alpha_{m_F} \approx -iM/{2\hbar t_3}$, which leads to a simplified form of interference pattern,
\begin {eqnarray}\label{eqn:interference}
&&|\psi _{+, m_F}(y)+\psi _{-,m_F}(y)|^{2}= \nonumber\\
&&{A_{m_F, {\rm const}}} [1+\frac{A_{m_F, {\rm osc }}}{A_{m_F, {\rm const}}} \cos \left( \kappa_{m_F} y+\phi_{m_F}\right)]
\end{eqnarray}

Here we have a constant part $ A _{m_F, {\rm const}}= \Gamma_{+,m_F}^2 + \Gamma_{-,m_F}^2$ and an oscillating part that leads to the interference fringes with the visibility $A _{m_F, {\rm osc}}/A _{m_F, {\rm const}}$, in which $A _{m_F, {\rm osc}}= 2\Gamma _{+,m_F}\Gamma _{-,m_F}$. $\kappa_{m_F} \thickapprox {Md/\hbar t_3}$ is the spatial fringe frequency; more details in Appendix~\ref{sec:Appendix2}. $\phi_{m_F}= \theta_{+, m_F}- \theta_{-, m_F}+(\gamma_{+,m_{F}}+\gamma_{-,m_{F}})*D/2$ is the global phase difference of the wave packets $\psi _{+, m_F}(y)$ and $\psi _{-, m_F}(y)$. This approximation helps us to see the physical factors that affect the fringe frequency, visibility, and phase more clearly.

\subsection{Fitting the interference patterns}
Due to experimental imperfectness, such as partial overlap between the two interfering wave packets, the finite temperature of the atomic cloud, and limited imaging resolution, we expect the visibility $V_{m_F}$ to be smaller than the theoretical value of ${A_{m_F, {\rm osc }}}/{A _{m_F, {\rm const}}}$[Eq.~\eqref{eqn:interference}]. In order to account for the real interference patterns, the visibility was included as a parameter and the atomic state at each stage of the interferometric process is assumed as a superposition of Gaussian wave packets~\cite{11Machluf2013}. So we use an empirical expression having the same form with Eq.~\eqref{eqn:interference} as the following
\begin {equation}\label{eqn:fitting}
\Lambda_{m_F}(y)=A_{m_F, {\rm const}} G_{m_F}(y) [1+V_{m_F} \cos (\kappa_{m_F} y+\phi_{m_F})]
\end{equation}
where $G_{m_F}(y)$ is a Gaussian function. The spatial fringe frequency $\kappa_{m_F}= 2\pi/\lambda_{m_F}$, in which $\lambda_{m_F}$ is the fringe periodicity. 

Equation.~\eqref{eqn:fitting} is used to fit the experimental data, shown as the solid curves in Figs.~\ref{fig:interference}($b1$)-~\ref{fig:interference}($b5$). From the fitting, we get that the visibility of the states $\left|1\right\rangle$, $\left|-1\right\rangle$, $\left|2\right\rangle$, and $\left|-2\right\rangle$ are about $0.6 \pm 0.1$, where the experimental parameters $t_1 = 210 \mu$s and $t_2 = 1300 \mu$s are chosen. Meanwhile the spatial fringe periodicities are about $26.5 \pm 1.0\mu m$. We observe almost no interference fringes in the $| 0\rangle$ species, for that the fraction of atom cloud $\psi _{-, 0}$ generated from the atom cloud $\psi _{-}$ after the MFP2 is tiny, which is consistent with the theoretical analysis in Appendix~\ref{sec:Appendix1}. Hence in the following we mainly study the interference patterns of the other four components.

If we take an average of these pictures in repeated experiments, the interference fringes actually disappear---the average of 15 shots washes out the interference fringes, as shown in Fig.~\ref{fig:interference}(c). This means that the phase repeatability is poor and the phase of each component in every experimental run is evenly distributed; more detailed measurements are given in Sec.~\ref{sec:phasestability} .

\section{Towards optimal visibility with time sequence design}

\subsection{Experimental optimization of the visibility of interference fringes}

The visibility is an important characteristic parameter of the interferometry. High-visibility interference patterns allow us to extract the phase, fringe frequency, and other interference information more accurately. However, the imperfect overlap will cause a drop in visibility. So we study the dependence of the interference fringes' visibility on the interfering wave packet separation $D$, where $D\thickapprox d+({a_c}{t_1}+v_r)t_3$ is tuned by the pulse interval $t_1$ after $t_3=26 ms$ TOF. In experiment, $t_1$ can be changed from zero microseconds to hundreds of microseconds to obtain different separation $D$ in Fig.~\ref{fig:procedure}(a).

\begin{figure}
\includegraphics[width=\linewidth]{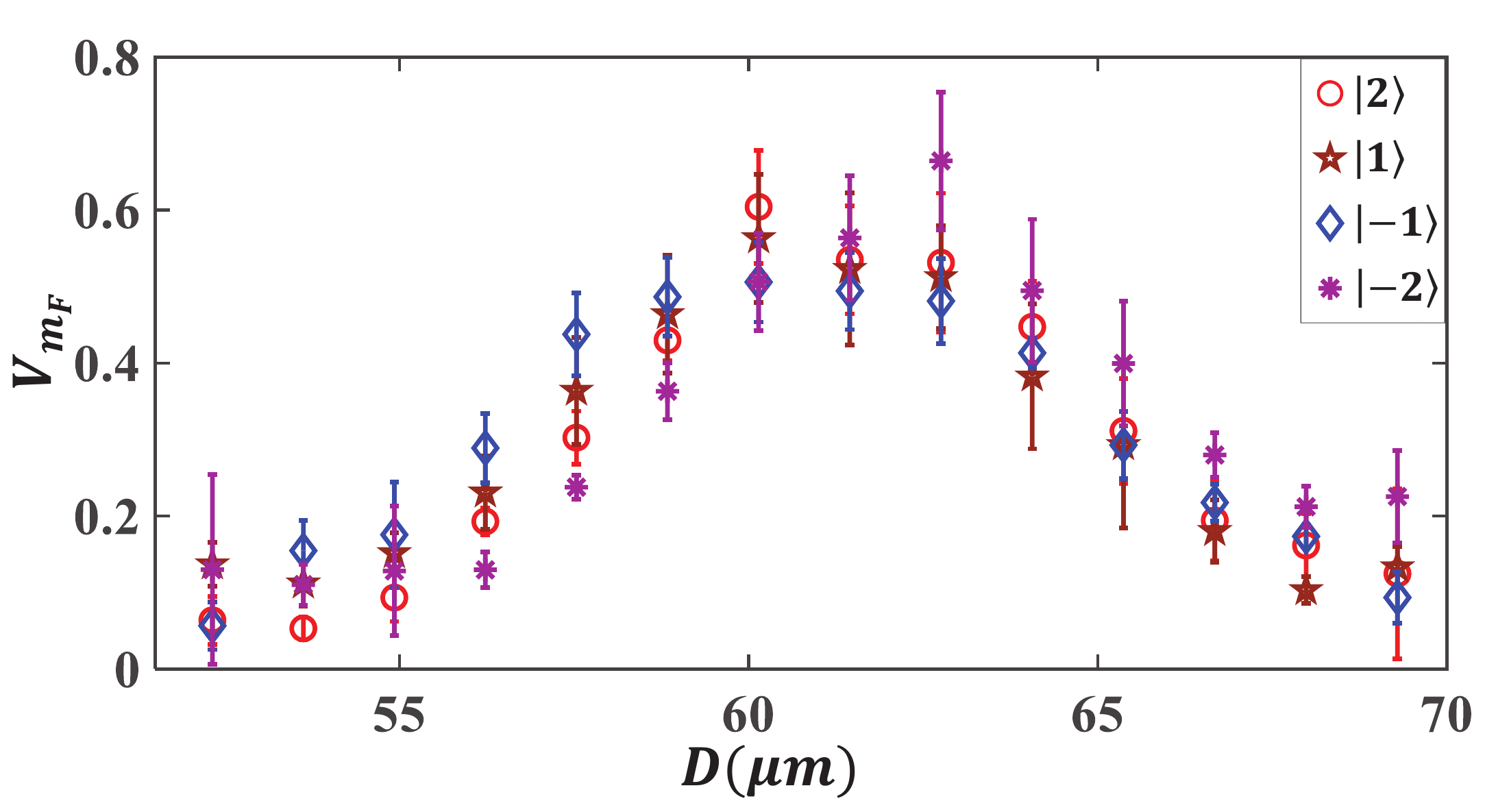}
\caption{(Color online) Measured visibility $V_{m_F}$ versus the separation $D$ for states $\left|2\right\rangle$ (circular), $\left|1\right\rangle$ (pentagram), $\left|-1\right\rangle$ (diamond), and $\left|-2\right\rangle$ (m shaped), respectively. Each experimental point is the average of three consecutive shots, where the experimental points of $V_{m_F}$ are obtained by fitting the patterns using Eq.~\eqref{eqn:fitting}.}
\label{fig:visibility}
\end{figure}

When $t_1$ is as small as tens of microseconds, there are almost no interference fringes observed, which can be attributed to large spatial distance between adjacent fringes. With $t_1$ increasing, the fringe visibility gradually increases. The separation $D$ can increase from $52\mu$m to $69\mu$m, which corresponds to the $t_1$ changing from $140\mu s$ to $270\mu s$ as shown in Fig.~\ref{fig:visibility}. By the experimental data, we observe a nonmonotonic dependence of visibility on separation $D$, with an optimal separation at $D_{\rm optimal} \approx 61\mu m$, which corresponds to $t_1 = 210\mu s$. The eventual decrease in the visibility at a larger $D$ is because of the decrease in the overlap of the interfering wave packets, which causes a smaller number of atoms in the interference region. So in order to gain high visibility interference fringes, a proper $t_1$ should be selected.

\subsection{Analysing the minor difference between the fringe frequencies of different components}

To characterize the property of the multicomponent interferometer, it is necessary to measure the difference between the fringe frequencies of different ${m_F}$ components. In order to consider the discrepancy, the form $\kappa_{m_F}$  without approximation should be used [seen in Eq.~(\ref{eqn:wavenumber1}]; it has the following form~\cite{13Simsarian2000}:
\begin {equation}\label{eqn:fit}
\kappa_{m_F} =\beta_{m_F}d + \zeta_{m_F}
\end{equation}
where the intercept $\zeta_{m_F}$ is a fitting parameter related to $\varepsilon_{m_F}$ and the final velocity difference between the interfering wave packets, as shown in Appendix~\ref{sec:Appendix2}.

\begin{figure}
\includegraphics[width=\linewidth]{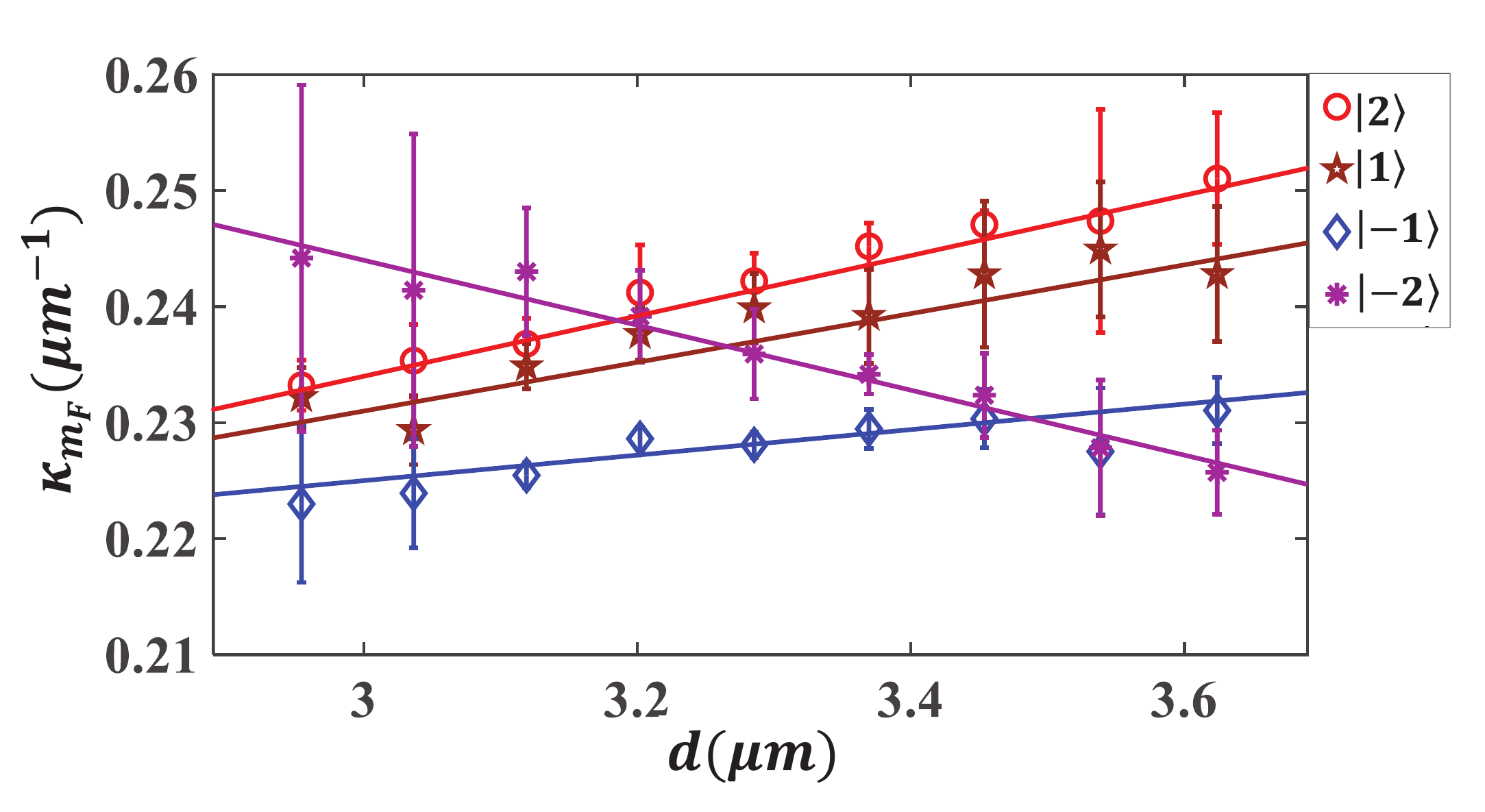}
\caption{(Color online) Spatial fringe frequency $\kappa_{m_F}$  vs the separation $d$ for states $\left|2\right\rangle$ (circular), $\left|1\right\rangle$ (pentagram), $\left|-1\right\rangle$ (diamond) and $\left|-2\right\rangle$ (m-shaped), respectively. Each experimental point is the average of three consecutive shots, where the experimental points of $\kappa_{m_F}$ are obtained by fitting the patterns using Eq.~\eqref{eqn:fitting}. The solid lines are linear fits using Eq.~(\ref{eqn:fit}).}
\label{fig:beta}
\end{figure}

We vary the distance $d$ by choosing the pulse interval $t_1$ from $170\mu$s to $250\mu$s. The experimental results for the different $\kappa_{m_F}$ are shown in Fig.~\ref{fig:beta}. For different hyperfine states, we obtain $\beta_{2}=0.026, \beta_{1}=0.021, \beta_{-1}=0.011, \beta_{-2}=-0.028$ by fitting the data with Eq.~(\ref{eqn:fit}). The intercepts are $\zeta_{2}=0.16, \zeta_{1}=0.17, \zeta_{-1}=0.19, \zeta_{-2}=0.32$. The results exhibit a weak $|m_F\rangle$ dependence of the coefficient $\beta_{m_F}$  of the quadratic phase of the wave function and the intercept term $\zeta_{m_F}$.

\section{The phase stability}
\label{sec:phasestability}
\begin{figure}
\includegraphics[width=\linewidth]{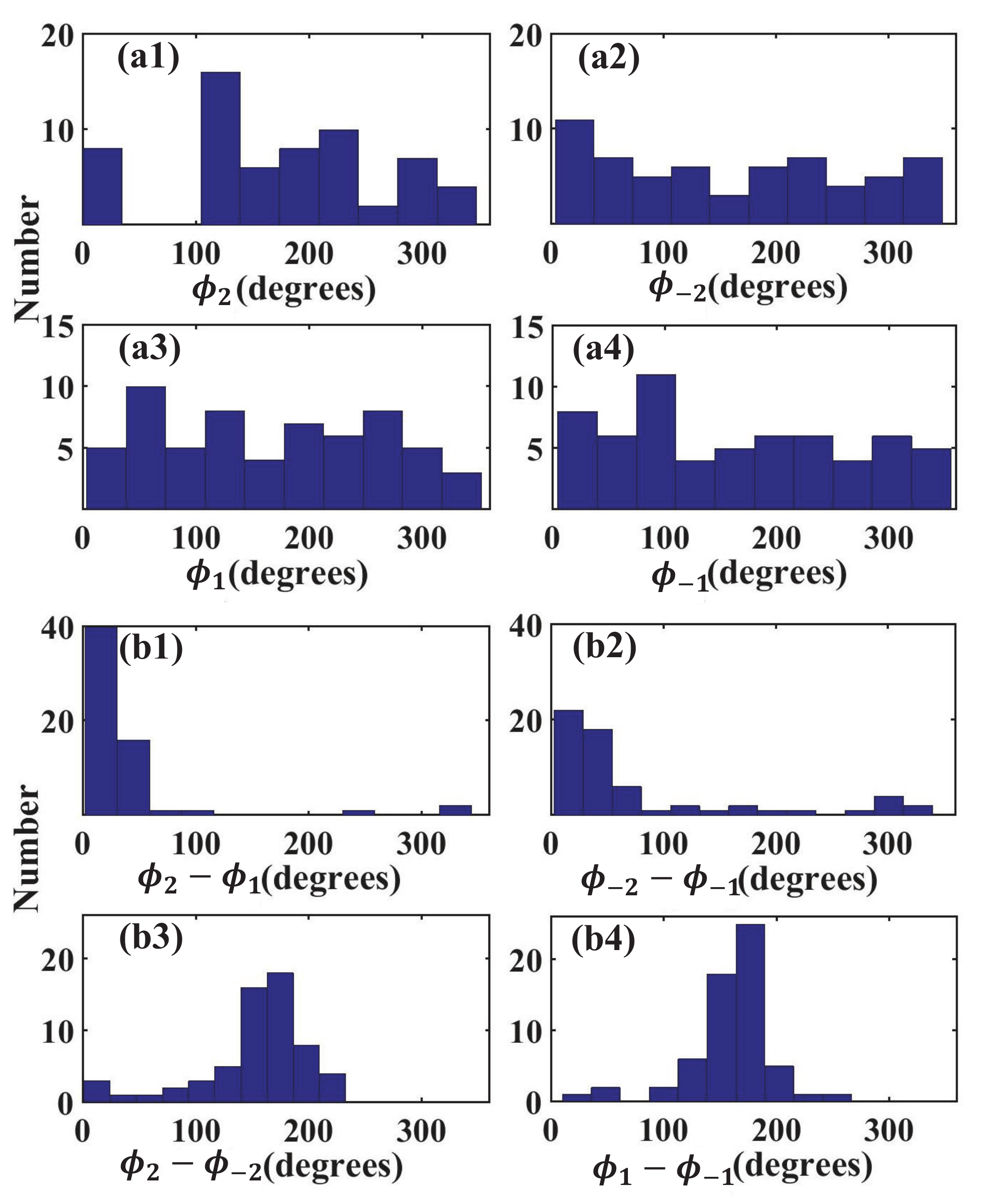}
\caption{(Color online) Histograms of phase distributions of 61 consecutive experimental shots, when the optical trap is turned off $t_0=3.8ms$ ahead of MFP1. (a1)-(a4) The distributions of phases ($\phi_2$, $\phi_{-2}$, $\phi_1$, and $\phi_{-1}$) are shown, respectively. These phases are basically evenly occupying all the phase spectrum. (b1)-(b4) The distributions of relative phases ($\phi_2- \phi_1$, $\phi_{-2}- \phi_{-1}$, $\phi_{2}- \phi_{-2}$, and $\phi_{1}- \phi_{-1}$) are shown, respectively. These relative phases show good repeatability.}
\label{fig:phase1}
\end{figure}

The phase $\phi_{m_F}$ is very important in the study of quantum coherence and also for precision measurements. By fitting the experimental data as shown in Fig.~\ref{fig:interference}($b$) using Eq.~(\ref{eqn:fitting}), we can get the phase information $\phi_{m_F}$ of interference fringes, which represents the phase difference between the two copies of our considered four spin components ($m_F=2,1,-1,-2$) after MFP2. To further study them, we give the experimental distributions of $\phi_{m_F}$ in continuously measured more than 61 experimental runs as shown in Fig.~\ref{fig:phase1}(a). The phase $\phi_{m_F}$ of each component is almost random, which means the phase difference between the two copies of each component is not fixed. This can be attributed to uncontrollable phase accumulation due to the fluctuations in the magnetic field during the adiabatic procedure of the MFP2~\cite{11Machluf2013}.

However, the relative phase across the four spin components---the change of the phase difference as we look at the four spin components---remains the same in different experimental runs. As shown in Fig.~\ref{fig:phase1}(b), the distributions of the relative phases (${\phi_2} - {\phi _1}$, ${\phi _{-2}} - {\phi _{-1}}$, ${\phi _2} - {\phi _{-2}}$, ${\phi _{1}} - {\phi _{-1}}$) are given from (b1) to (b4), respectively. The main values for the ${\phi _2} - {\phi _1}$ and ${\phi _{-2}} - {\phi _{-1}}$ are concentrated at about zero degree, while the latter two are concentrated at about 180 degrees, where $t_0=3.8ms$, $t_1 = 210\mu s$ and $t_2 = 1300\mu s$.  These relative phases are stable because the four spin components in each of the two copies of the atomic wave packets have to share the same phase in our experiment.

\begin{figure}
\includegraphics[width=\linewidth]{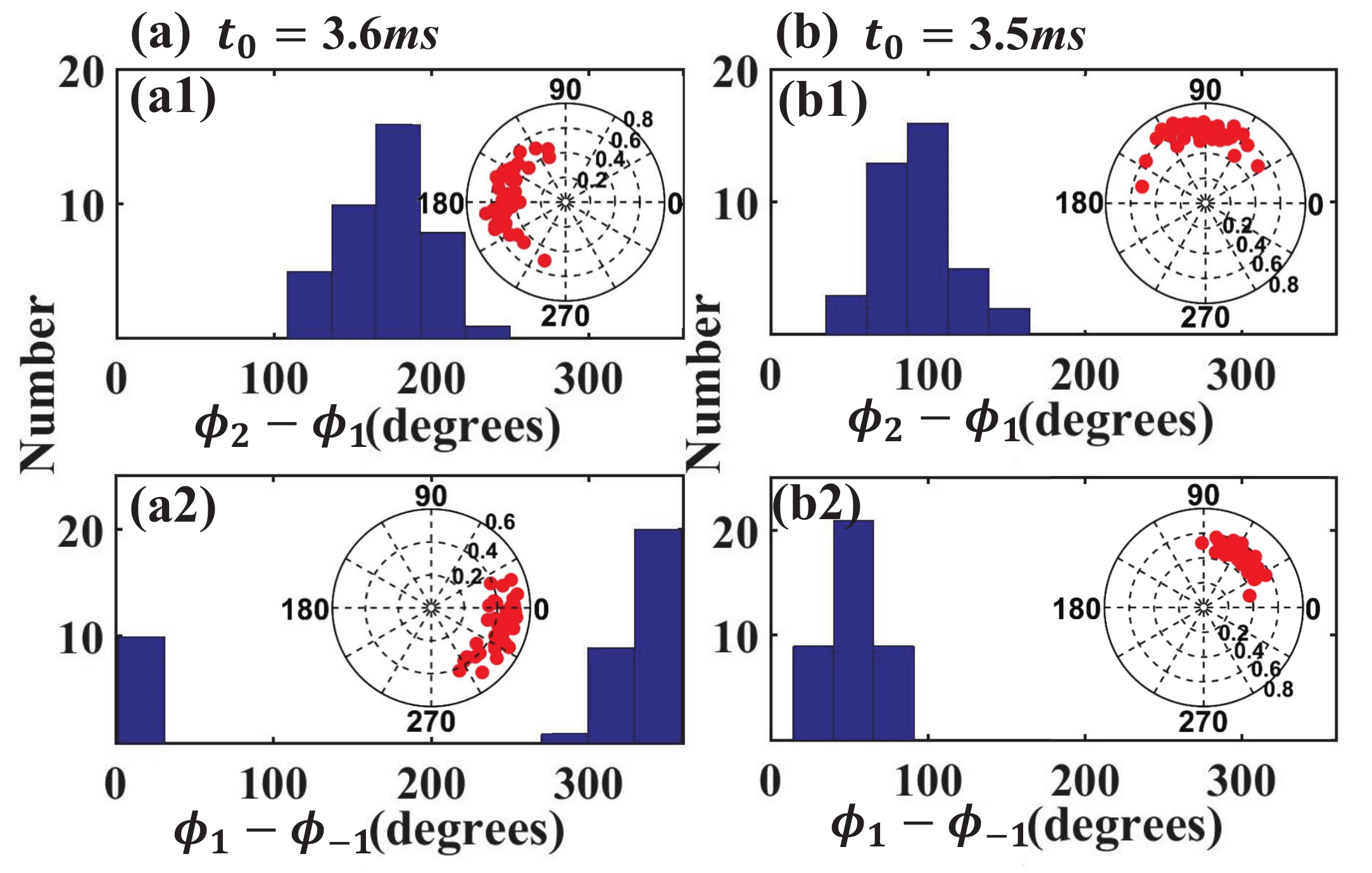}
\caption{(Color online) Relative phase distributions of 41 consecutive experimental shots with different $t_0$ in (a) and (b). (a) The distributions of relative phases $\phi_2- \phi_1$ and $\phi_1- \phi_{-1}$ are shown in (a1) and (a2), respectively, when $t_0=3.6ms$. (b)  When $t_0=3.5ms$, the distributions of relative phases $\phi_2- \phi_1$ and $\phi_1- \phi_{-1}$ are shown in (b1) and (b2), respectively. The polar plots of relative phase vs visibility (shown as angle vs radius) are shown as these insets, respectively, where the value of visibility is an average of the visibility of corresponding two components.}
\label{fig:phase2}
\end{figure}

The relative phase can be controlled by changing the time $t_0$, which mainly leads to a minor difference (about a few microns) in BEC position in quadrupole magnetic trap and a tiny change of the phase profile of the BEC~\cite{13Simsarian2000,14Castin1996}. If we change the interval time to be $t_0=3.6ms$, the distributions of the relative phases ${\phi _2} - {\phi _1}$ and ${\phi _{1}} - {\phi _{-1}}$ are given in Figs.~\ref{fig:phase2}(a1) and~\ref{fig:phase2}(a2). They are roughly concentrated at 180 degrees and 350 degrees, respectively. When $t_0=3.5ms$, these relative phases are changed to focus at about 90 degrees and 60 degrees, as shown in Figs.~\ref{fig:phase2}(b1) and ~\ref{fig:phase2}(b2), respectively. This means we can obtain the steady phase difference between different components by adjusting the time that the optical trap is turned off in advance of MFP1. It makes the multicomponent interferometer a potential application for precision measurement, such as measuring the small changes in the spin-dependent external potential, or leading to multi-pointer interferometric clocks~\cite{margalit2015self}. 

\section{Conclusion}
In this paper we report the realization of a parallel multicomponent interferometer by integrating experimental techniques of spinor condensate, Majorana transition, and Stern-Gerlach momentum splitting. We demonstrate four atomic spatial interference fringes at the same time, using four out of five spinor components in a spin-2 system after the TOF. By controlling magnetic fields, we control the spin projection in different sublevels with Majorana transitions to achieve a multispin double slit interference. The coherent Stern-Gerlach splitting and TOF analysis are used to characterize the magnetic order in a spinor gas, which helps us require knowledge of the fringe frequency, visibility, and phase of each interference fringe. By tuning the separation of the interfering wave packets, the visibility of different component fringes of the interferometer can be optimized to be the best at the same time. The relative phase across the different hyperfine states is found to be stable in repeated experiments, despite that the phase of each individual interference fringe appears to be not repeatable. The implemented noise-resilient parallel interferometer together with the developed techniques would potentially lead to multi-pointer interferometric clocks~\cite{margalit2015self}.

\medskip
\section{Acknowledgements}
We thank Heng Fan, Baoguo Yang, Xinhao Zou, Shuyang Cao, Ji Li and Peng Zhang for helpful discussion. This work is supported by the National Key Research and Development Program of China (Grants No. 2016YFA0301501 and No. 2017YFA030420) and the National Natural Science Foundation of China (Grants No. 61727819, No. 91736208, and No. 117740067).

\appendix
\addcontentsline{toc}{section}{Appendices}\markboth{APPENDICES}{}
\begin{subappendices}

\section{The theoretical explanation of the spin-projection achieved by the operation MFP2}\label{sec:Appendix1}
In order to understand why the $\left| 0\right\rangle$ state is basically free of fringes, we give a theoretical explanation as followed through the analysis of the Majorana transition process. We have the initial prepared condensate in state $\left|F,m_F \right>$, where the atomic spin $F=2$ and submagnetic level $m_F=2$. After the Majorana transition and the Stern-Gerlach process, the condensate can be expressed as
\begin{eqnarray}\label{eqn:Psi}
\left| \Psi \right\rangle = \sum\limits_{m_F = -2}^2 {\left| \psi^\prime_{m_F} \right\rangle \otimes \left| {{p_{m_F}}} \right\rangle }
\end{eqnarray}
where ${p_{m_F}}$ is the momentum of the submagnetic state $\left| \psi^\prime_{m_F} \right\rangle$ and $\left| \psi^\prime_{m_F} \right\rangle = \Gamma_{m_F} {e^{i{\varphi _{m_F}}}}\left| m_F \right\rangle$, in which $\Gamma_{m_F}$ represents the amplitude of the wave function; $\varphi _{m_F}$ is the phase of the $\left| m_F \right\rangle$-state wave function.

By neglecting the part of momentum, we have $\left| \Psi^\prime \right\rangle = \sum\limits_{m_F = -2}^2 {\left| {{\psi^\prime _{m_F}}} \right\rangle } \doteq {\left( {\begin{array}{*{20}{c}} {\Gamma_{2} {e^{i{\varphi _2}}}}\cdots {\Gamma_{-2} {e^{i{\varphi _{ - 2}}}}}
\end{array}} \right)^{\text{T}}}$. Suppose that, after the operation MFP1 in experiment, the initial state is populated into only two states of all the submagnetic states; they are
$\left| 2 \right\rangle \doteq {\left( {\begin{array}{*{20}{c}}
	1&0&0&0&0
	\end{array}} \right)^{\text{T}}}$ and $\left| 1 \right\rangle \doteq {\left( {\begin{array}{*{20}{c}}
	0&1&0&0&0
	\end{array}} \right)^{\text{T}}}$.
Then $\left| \Psi^\prime \right\rangle$ reduces to
\begin{eqnarray}
\textstyle
\left| \Psi^\prime_0 \right\rangle \doteq {\left( {\begin{array}{*{20}{c}}
	{\Gamma_{2} {e^{i{\varphi _2}}}}
	\  {\Gamma_{1} {e^{i{\varphi _{1}}}}
		\  0\ 0\ 0
	}
	\end{array}} \right)^{\text{T}}}
\end{eqnarray}

The Majorana formula has been derived from group theory~\cite{51Angular}. For a multilevel system with a total angular momentum $F$, it is
\begin{eqnarray}\label{eqn:D_elements}
{{\hat U}_{{m_F}',m_F}}\left( {a,b} \right) =\sum\limits_n {{{\left( { - 1} \right)}^n}\xi} \times \chi
\end{eqnarray}
where ${\hat U}^\dag$ is a rotation operator and ${{\hat U}^\dag }\hat U = 1$. $a$ and $b$ are both related to the rotation angles, and ${\left| a \right|^2} + {\left| b \right|^2} = 1$.
$\xi=\frac{{\sqrt {\left( {F - m_F} \right)!\left( {F + m_F} \right)!\left( {F - {m_F}'} \right)!\left( {F + {m_F}'} \right)!} }}{{\left( {F + {m_F}' - n} \right)!\left( {F - m_F - n} \right)!n!\left( {n + m_F - {m_F}'} \right)!}}$ and $\chi={{a^{F + {m_F}' - n}}{{\left( {{a^*}} \right)}^{F - m_F - n}}{b^n}{{\left( {{b^*}} \right)}^{n + m_F - {m_F}'}}}$;  the value $n$ in the summation contains all the integers making the four factorials in the denominator reasonable.  According to the effect of operation MFP2 in experiment, we can take $a = b $; the concrete form is
\begin{eqnarray}\label{eqn:D1}
\hat U\left({\frac{1}{{\sqrt 2 }},\frac{1}{{\sqrt 2 }}} \right) = \frac{1}{4}\left( {\begin{array}{*{20}{c}}
1&{ - 2}&{\sqrt 6 }&{ - 2}&1 \\
2&{ - 2}&0&2&{ - 2} \\
{\sqrt 6 }&0&{ - 2}&0&{\sqrt 6 } \\
2&2&0&{ - 2}&{ - 2} \\
1&2&{\sqrt 6 }&2&1
\end{array}} \right)
\end{eqnarray}
$\\$

After the MFP2 in experiment, the wave function is
\begin{eqnarray}
\hat U \left|\Psi^\prime_0 \right\rangle
\doteq \frac{1}{4} {\left( {\begin{array}{*{20}{c}}
	{\Gamma_{2} {e^{i{\varphi _2}}}-2\Gamma_{1} {e^{i{\varphi _{1}}}}} \\
	%  \ \\
	{2\Gamma_{2} {e^{i{\varphi _2}}}-2\Gamma_{1} {e^{i{\varphi _{1}}}}}\\
	%  \ \\
	{\sqrt{6} \Gamma_{2} {e^{i{\varphi _2}}}} \\
	% \ \\
	{2\Gamma_{2} {e^{i{\varphi _2}}}+2\Gamma_{1} {e^{i{\varphi _{1}}}}}\\
	%  \ \\
	{\Gamma_{2} {e^{i{\varphi _2}}}+2\Gamma_{1} {e^{i{\varphi _{1}}}}}
	\end{array}}\right)}
\end{eqnarray}
$ \\$

By calculating $\left|\left<{m_F}\right|\hat U \left|\Psi^\prime_0  \right\rangle\right|^2 $, we can know the density distribution of all the $\left| m_F \right\rangle$ states. Since there is only one condensate copy in the $\left|0\right\rangle$ species after the MFP2, $\left|\left<0\right|\hat U \left|\Psi^\prime_0 \right\rangle\right|^2 = 3\Gamma_{2}^2/8$ just contains the constant part without an oscillating part. So there are almost no interference fringes in the $| 0\rangle$ species, as shown in Fig.~\ref{fig:interference}($b5$).

\section{Analysis of the force applied on $\left| m_F\right\rangle$-state wave packet in the GMF}\label{sec:Appendix0}
In the quadrupole magnetic trap, the force that the $|m_F\rangle$-state wave packet suffered from is $\mu _{b}m_{F}g_{F}B_y^{'}$ along the $y$ direction, which is supplied by the gradient magnetic field.  $\mu _{b}$ is the Bohr magneton; $g_F$ is the Landau g-factor of an atom with total angular momentum $F$. Since the BEC in $|2,2\rangle$ state is prepared in the hybrid trap and located about $158 \mu$m below the center of the quadrupole magnetic trap, we have $2\mu _{b}g_{F}B_y^{'}+f_{opt}= Mg$, where $g$ is the gravity acceleration. The force $f_{opt}$ supplied by the optical trap is very small, which is about five percent of the force supplied by the gradient magnetic field. So when the optical trap is shut off, the resultant force applied on the $\left| m_F\right\rangle$-state wave packet is $f_{m_F} = (2-m_F)\mu _{b}g_{F}B_y^{'}+f_{opt}$. The force $f_{opt}$ is independent of the different submagnetic states, {so the relative acceleration between adjacent submagnetic states in the Stern-Gerlach process is $a_c=(f_{m_F} -f_{m_F-1})/M= \mu _{b}g_{F}B_y^{'}/M$.}

\section{The expression of spatial fringe frequency $\kappa_{m_F}$ }\label{sec:Appendix2}
When the long TOF approximation isn't considered, we have the imaginary part $\beta_{m_F}=-2\text{Im}[\alpha _{m_{F}}]=M/[\hbar(\varepsilon_{m_F}+t_3)]$. In the process of calculating $|\psi _{+, m_F}(y)+\psi _{-,m_F}(y)|^{2}$, we have
\begin{eqnarray}\label{eqn:wavenumber}
\kappa_{m_F}&& = \beta_{m_F}D-\frac{M(v_{-, m_F}-v_{+, m_F})}{\hbar}
\end{eqnarray}

Using the formulas $D=d+\delta v|_{T_2} t_3$ and $v_{-, m_F}-v_{+, m_F}= \delta v|_{T_2}$, which are mentioned above, in the Eq.~\eqref{eqn:wavenumber}, we can derive the following relation: 
\begin{eqnarray}\label{eqn:wavenumber1}
\kappa_{m_F}=&&\frac{Md}{\hbar(\varepsilon_{m_F}+t_3)}+\zeta_{m_F} 
%\nonumber\\=&&\beta_{m_F} d+\zeta_{m_F}
\end{eqnarray}
where $\zeta_{m_F}= {M(v_{+, m_F}-v_{-, m_F})\varepsilon_{m_F}}/{\hbar(\varepsilon_{m_F}+{t_3})}$ represents the item of interference fringes caused by the small item of $\varepsilon_{m_F}$ and the relative velocity between the interfering wave packets.
Under the long TOF limit( $t_3\gg \varepsilon_{m_F}$)~\cite{11Machluf2013} and considering the final velocity difference between the interfering wave packets is smaller than the expansion velocity of each one of them [which means $d\gg\varepsilon_{m_F}(v_{+, m_F}-v_{-, m_F})$], Eq.~\eqref{eqn:wavenumber1} reduces to
\begin{eqnarray}\label{eqn:wavenumber2}	
\kappa_{m_F}\thickapprox Md/\hbar t_3
\end{eqnarray}

The $\kappa_{m_F}$ in Eq.~(\ref{eqn:wavenumber2}) is an approximate form of Eq.~(\ref{eqn:wavenumber1}) in the long TOF limit, which is  independent of different spin states. However, Eq.~(\ref{eqn:wavenumber1}) is more accurate for fitting the fringe frequency and can give the discrepancy of fringe frequencies between different submagnetic states.

\end{subappendices}

\bibliographystyle{apsrev}
\bibliography{citelist}

\end{document}